# Managing Activities at the Lunar Poles for Science[*]

Ian A. Crawford[1], Parvathy Prem[2], Carlé Pieters[3], and Mahesh Anand[4]

[1] Birkbeck College, University of London, London, UK; i.crawford@bbk.ac.uk
[2] Johns Hopkins Applied Physics Laboratory in Laurel, Maryland, USA
[3] Brown University in Providence, RI, USA
[4] The Open University, Milton Keynes, UK

## 1. Introduction

As noted by the US National Research Council (NRC) Report on the Scientific Context for Exploration of the Moon [1], the lunar poles are special environments deserving careful scientific study as the international community begins detailed exploration in the years ahead. Although little direct information about these special areas is available, it is possible that they are fragile environments that, once disturbed, cannot be restored to their natural state. It follows that these environments merit special protection from disruptive interference. To an extent, this has been recognized in recent changes to COSPAR's Planetary Protection Policy (PPP) [2], but there are reasons to believe that these protections may not be sufficiently strong.

## 2. Scientific Background

Scientific interest in the lunar poles is mainly focused on volatiles that are likely to be preserved as ices in Permanently Shadowed Regions (PSRs). There are at least two reasons why volatiles at the lunar poles are scientifically valuable:

- Lunar polar volatiles may provide a record of volatiles and pre-biotic organic materials delivered to the Earth-Moon system (and thus to Earth) from elsewhere in the Solar System (and possibly beyond) [1].

- Lunar polar ices irradiated by cosmic rays may provide a natural laboratory for the synthesis of organic molecules relevant to understanding comparable processes on the surfaces of icy bodies in the outer Solar System and on interstellar dust grains. Understanding these processes is of great astrobiological interest, and the lunar poles are probably the most easily accessible locations where they can be studied [3,4].

In addition to enabling the study of trapped volatiles, certain aspects of the lunar polar environments, especially PSRs, may also be uniquely suitable as locations for some types of astronomical observations (e.g., infrared [5] and gravitational wave [6,7] observations). Such applications will rely on the maintenance of a high vacuum, minimal vibrations, and (to the extent possible) a dust-free environment.

---

For all these reasons, it seems desirable to preserve lunar polar environments from disturbance by extraneous volatiles (e.g., from rocket exhausts), organic materials, and the generation and widespread dispersal of lunar dust. However, there is a potential tension between these scientific interests and plans for the development of scientific and commercial infrastructures at the lunar poles. For example, lunar polar ices may be a valuable source of water for in situ resource utilization (ISRU) to support further scientific exploration of the Moon, as well as possibly to kick-start a cis-lunar economy [8]. Science stands to benefit from these activities [9,10], but the extent to which they may negatively modify pristine polar environments remains largely unknown and is an important topic of on-going study [11-13].

## 3. Preserving individual PSRs

One way to learn about the extent of contamination of PSRs caused by future exploration activities, and perhaps help mitigate the consequences, would be to designate specific PSRs for special protection. Fig. 1 shows the distribution of PSRs around both lunar poles as mapped by Mazarico et al. [14]. Although the areas of permanent shadow are generally larger around the South Pole, it can be seen that numerous PSRs exist within 10° of latitude (~300 km) of both poles. By agreeing to place a moratorium on surface operations in some of these PSRs, apart from the placement of passive monitoring instruments within a subset of them, the extent of contamination arising from more active vigorous human or robotic activities around the lunar poles could be assessed.

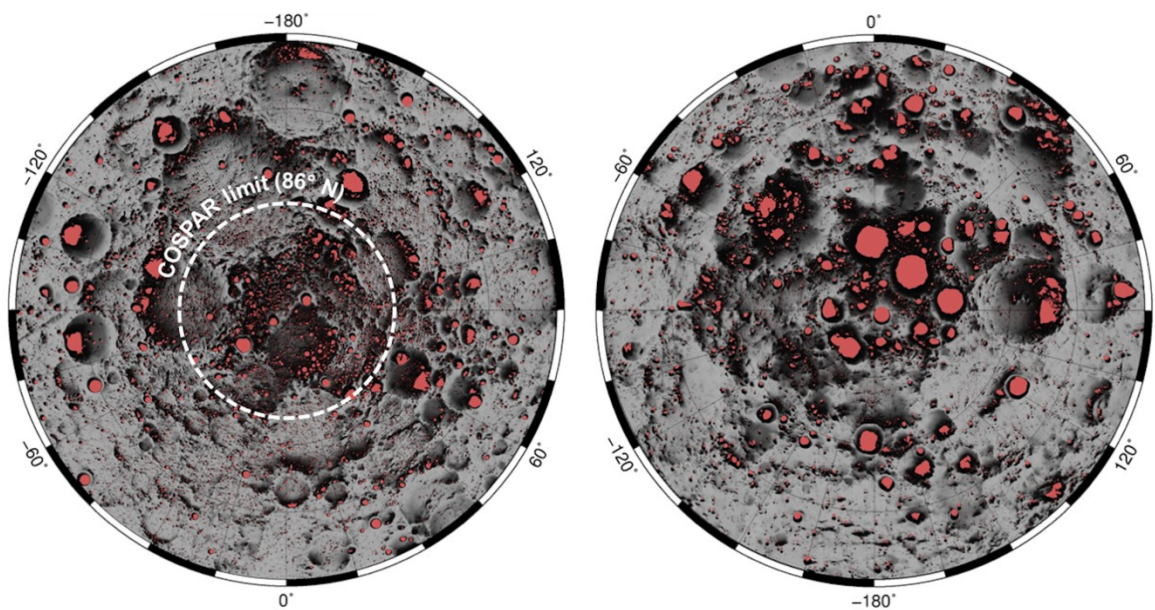

Fig. 1. Locations of permanent shadow (marked in red) overlaid on average solar illumination (grayscale) within 10° latitude of the lunar North Pole (left) and South Pole (right). The northern latitude limit for COSPAR PPP Category II(b), 86°N, is indicated; the corresponding southern limit, 79°S, lies outside the area shown. Data from https://pgda.gsfc.nasa.gov/products/69 (see Mazarico et al. [14]).

Ideally, the PSRs identified for special protection and monitoring would be internationally agreed upon (possibly via a revised COSPAR Planetary Protection Policy, see below), and chosen to provide a geographically representative distribution at a range of distances from sites of exploration and commercial activities. The recent US National Academies 'Report by the Committee on Planetary Protection: Planetary Protection for the Study of Lunar Volatiles' [15] alludes to the possibility of establishing PSR protected areas (p. 6), but falls short of making any specific recommendations in this respect.

The same study presents several reasons why significant contamination of PSRs by spacecraft operations may be unlikely, but notes (p. 15) that

> *[g]iven the incomplete knowledge of volatile transport across the lunar surface, it would be wise to exercise caution when interacting with these special deposits so that their value can be assessed and tapped.*

Placing a moratorium on operations within, or close to, an internationally agreed set of PSRs, while environmental impact assessments are made in non-protected PSRs, would go some way towards satisfying this recommendation. If, after a suitable period (to be determined, but perhaps several decades), environmental monitoring indicates that significant contamination from human activity does not occur, the protected status could be revised.

## 4. Preserving the North Pole

On the other hand, we should consider what to do if contamination instead turns out to be widespread, perhaps extending for hundreds or thousands of kilometres from sites of exploration or commercial activities. If such activities were to be initiated at both poles before we discover that contamination can be widespread, the opportunity to preserve pristine PSRs may be lost. It is therefore fortunate that the Moon has two poles, whereas to date most exploration activities proposed by both government space agencies and commercial operators are focused on areas around the South Pole [16].

Thus, in principle, the entire North Polar region (say north of 80°N) might be set aside as a protected area, effectively out of bounds to surface exploration, at least until we understand more about how human and robotic operations have affected scientifically valuable locations and deposits around the South Pole. Some volatiles released by human activities close to the South Pole may migrate to PSRs around the North Pole, but the study by Prem et al. [12] suggests that this is likely to be an order of magnitude less than contamination of southern PSRs by the same activity. In any case, the North Pole is as far from the South Pole as it is possible to get on the lunar surface, so protecting it will be the best that we can do.

Realistically, such a moratorium might have to last for several decades, but this would depend on the pace of exploration and commercial activities around the South Pole and on-going assessments of their environmental consequences. A more permanent possibility, should this be deemed desirable, might be to designate the lunar North Pole as a 'Planetary Park' along the lines suggested by Cockell and Horneck [17]. Either of these suggestions would ensure much stronger protection for PSRs at the lunar North Pole than currently mandated by the COSPAR Planetary Protection Policy [2], which, we suggest, is not sufficient to exclude the possibility of serious contamination of lunar polar environments.

## 5. Note on the COSPAR Planetary Protection Policy (PPP)

The Moon falls within Category II of the PPP, covering "missions to those target bodies where there is significant interest relative to the process of chemical evolution and the origin of life, but where there is only a remote chance that contamination carried by a spacecraft could compromise future investigations" [2]. However, Category II does not place any limits on what may be taken to the Moon and requires only basic documentation.

Recently, the COSPAR PPP has been revised to acknowledge the scientific importance of PSRs by splitting Category II for lunar landing missions into IIa and IIb [2]. Category IIa, which applies to most of the lunar surface, requires that spacecraft operators provide an inventory of organic materials that may be released into the lunar environment by a spacecraft propulsion system. Category IIb applies to:

> *[a]ll missions to the surface of the Moon whose nominal profile access Permanently Shadowed Regions (PSRs) and the lunar poles, in particular latitudes south of 79°S and north of 86°N.*

In addition to the requirements of Category IIa, it adds a requirement that operators provide "a listing of all organic materials carried by a spacecraft which are present in a total mass greater than 1 kg."

The definition of Category IIb is rather confusing because there are many northern PSRs south of 86°N (see Fig. 1). Note that the latitude limits adopted for Category IIb leave the North Pole *less* protected than the South Pole, because a spacecraft would be free to land 120 km from the North Pole without providing an organic inventory, whereas the corresponding distance from the South Pole would be 330 km. The reasons for this asymmetry in the COSPAR PPP latitude limits are not explicitly stated in the document and it would be helpful if they could be clarified [18].

Moreover, and regardless of the choice of Category IIb latitude limits, the COSPAR PPP does not prevent the transport of organic materials to anywhere on the lunar surface (provided that the appropriate documentation is provided), and makes no mention of non-organic contamination or other disturbances (e.g. mobilized dust). Thus, it cannot be relied upon to prevent scientifically deleterious interference with the lunar polar environment. This provides a further argument for characterising the effects of spacecraft operations on the environment at the South Pole, and developing appropriate mitigation measures, before embarking on comparable activities at the North Pole. Indeed, experience gained at the South Pole may lead to additional protocols to protect the North Polar environment, and thus it will be important that the North Pole be left undisturbed while this experience is accumulated and new policies are developed.

## 6. Extending protection to other areas of the Moon (and elsewhere)

Establishing criteria to protect special areas may also be desirable in other regions of the Moon. Examples include the farside, where preserving or accommodating radio quietness will be important for radio astronomy [19], the unique environments likely to exist within sub-surface lava tubes [20], and other environmentally and geologically unique locations. This is not to suggest that all such localities be placed out of bounds for exploration, but rather that an internationally agreed sub-set of locations be identified and protected from interference while the effects of *in situ* exploration activities at comparable sites elsewhere on the Moon are assessed. Similar arguments might be advanced for the protection of cultural heritage sites on the Moon, as recommended in Section 9 of the recently formulated Artemis Accords [21].

Similar policies might profitably be adopted elsewhere in the Solar System. For example, the current strict PPP criteria for spacecraft visiting 'Special Regions' on Mars (Category IVc) is arguably inhibiting the search for life on that planet [22], whereas preventing *in situ* investigations of some regions currently placed in this category, while relaxing the constraints on visiting others, might strike a better balance between planetary protection and scientific investigations. We note, in passing, that the desirability of reaching international agreement on sites of special scientific interest (SSSI) meriting special protection was explicitly identified in Article 7(3) of the Moon Agreement of 1979 [23]. Notwithstanding the political reality that the Moon Agreement is currently undersubscribed by spacefaring nations, the identification of SSSI on the Moon and other planetary bodies appears to be a sensible recommendation, and could perhaps build on the concept of Antarctic Specially Protected Areas (ASPAs) as defined in Annex V of the Environmental Protocol to the Antarctic Treaty [24].

## 7. Conclusions

The lunar poles are unique environments of both great scientific and, increasingly, commercial interest. Consequently, a tension exists between the twin objectives of:

(a) Exploring the lunar poles for both scientific and commercial purposes, and utilising polar volatiles for ISRU and, ultimately, supporting a lunar economy; and

(b) Minimising the environmental impacts on the Moon's polar regions so as to preserve them for future scientific investigations.

We have argued here that the best compromise between these equally valuable, but not fully compatible, objectives would be to restrict scientific and commercial activities to the lunar South Pole, while placing a moratorium on activities at the North Pole until the full consequences of human activities at the South Pole are fully understood and mitigation protocols established. Depending on the pace at which lunar exploration proceeds, such a moratorium might last for several decades in order to properly assess the effects of exploration and commercial activities in regions surrounding the South Pole. An attractive longer-term possibility might be to consider designating the lunar North Polar region as a (possibly temporary) 'Planetary Park' [17]. Similar protected status might also be desirable for other unique lunar environments, and, by extension, other scientifically important localities elsewhere in the Solar System.